\pdfoutput=1

\documentclass[journal, pdftex]{IEEEtran}

\IEEEoverridecommandlockouts

\usepackage{tikz}
\usepackage{textcomp}
\usepackage{hyperref}
\usepackage{lipsum}

\usepackage{draftwatermark}
\SetWatermarkText{\textsc{Preprint}}
\SetWatermarkScale{.5}
\SetWatermarkColor[gray]{.83}

\newcommand\copyrighttext{%
  \footnotesize \textcopyright 2021 IEEE. Personal use of this material is permitted. Permission from IEEE must be obtained for all other uses, in any current or future media, including reprinting/republishing this material for advertising or promotional purposes, creating new collective works, for resale or redistribution to servers or lists, or reuse of any copyrighted component of this work in other works.
  DOI: \href{https://doi.org/10.1109/TE.2021.3123889}{10.1109/TE.2021.3123889}}
\newcommand\copyrightnotice{%
\begin{tikzpicture}[remember picture,overlay]
\node[anchor=south,yshift=7pt] at (current page.south) {\fbox{\parbox{\dimexpr\textwidth-\fboxsep-\fboxrule\relax}{\copyrighttext}}};
\end{tikzpicture}%
}

\usepackage{cite}
\usepackage{url}
\usepackage{hyperref}
\usepackage{mdframed}

\ifCLASSINFOpdf
  \DeclareGraphicsExtensions{.pdf}
\else
\fi

\usepackage{makecell}

\newcommand{\al}[1]{\thead[l]{#1 \\ \newline}}
\newcommand{\ar}[1]{\thead[r]{#1 \\ \newline}}

\newcommand{\eg}{\textit{e}.\textit{g}., }
\newcommand{\ie}{\textit{i}.\textit{e}., }
\newcommand{\etal}{\textit{et al}.}

\newmdtheoremenv[%
  outerlinewidth=4pt,
  leftmargin=0em,
  rightmargin=0em,
  innertopmargin=4pt,
  splittopskip=\topskip,
  skipbelow=\baselineskip,
  skipabove=.5\baselineskip]{implication}{Implication}

\begin{document}

\title{Software Engineering Education Knowledge \\ versus Industrial Needs}

\author{Georgios~Liargkovas,
        Angeliki~Papadopoulou,
        Zoe~Kotti,
        and~Diomidis~Spinellis,~\IEEEmembership{Senior~Member,~IEEE}
\thanks{The authors are with the Department of Management Science and Technology,
Athens University of Economics and Business, Athens, Greece
(e-mail: \{gliargovas,t8180094,zoekotti,dds\}@aueb.gr).}
}



%


\maketitle
\copyrightnotice

\begin{abstract}
\emph{Contribution:}
Determine and analyze the gap between
software practitioners' education outlined in
the 2014 IEEE/ACM Software Engineering Education Knowledge (SEEK)
and industrial needs pointed by
Wikipedia articles referenced in Stack Overflow (SO) posts.

\emph{Background:}
Previous work has uncovered deficiencies in
the coverage of computer fundamentals,
people skills,
software processes, and
human-computer interaction, 
suggesting rebalancing.

\emph{Research Questions:}
1) To what extent are developers' needs,
in terms of Wikipedia articles referenced in SO posts,
covered by the SEEK knowledge units?
2) How does the popularity of Wikipedia articles relate to their SEEK coverage?
3) What areas of computing knowledge can be better covered by the SEEK knowledge units?
4) Why are Wikipedia articles covered by the SEEK knowledge units
cited on SO?

\emph{Methodology:}
Wikipedia articles were systematically collected from SO posts.
The most cited were manually mapped to the SEEK knowledge units,
assessed according to their degree of coverage.
Articles insufficiently covered by the SEEK were classified by hand
using the 2012 ACM Computing Classification System.
A sample of posts referencing sufficiently covered articles was manually analyzed.
A survey was conducted on software practitioners 
to validate the study findings.

\emph{Findings:}
SEEK appears to cover sufficiently
computer science fundamentals,
software design and mathematical concepts,
but less so areas like
the World Wide Web,
software engineering components, and computer graphics.
Developers seek advice, best practices and explanations about software topics,
and code review assistance.
Future SEEK models and the computing education
could dive deeper in information systems, design, testing, security,
and soft skills.
\end{abstract}

\begin{IEEEkeywords}
Software engineering education, curriculum models, Stack Overflow, Wikipedia.
\end{IEEEkeywords}

%
\IEEEpeerreviewmaketitle

\section{Introduction}
\label{sec:intro}
\IEEEPARstart{O}{ver} the last decades
a lot of research has dealt with the lack of university curricula
focused on software engineering (SE).
In 1999,
Hilburn and Bagert~\cite{839269} stressed the need
for creating a contemporary SE curriculum model,
characterizing such a design as challenging,
due to the field's immaturity at the time.
For this purpose,
they proposed a model for designing SE curricula.
Earlier, Hilburn \etal~\cite{10.1007/3-540-58951-1_94}
had examined the integration
of SE principles
into undergraduate computer science programs,
proposing a structure
integrating eight fundamental software concepts 
in computing curricula with SE tracks.

SE is closely related with the computer science 
and computer engineering disciplines.
According to the latest IEEE/ACM Curriculum Guidelines
for Undergraduate Degree Programs
in Software Engineering (SE2014)~\cite{ABHO14},
SE curricula are usually developed by Departments 
of these two disciplines,
and include a combination of
computer science,
engineering,
and management science courses---this
distinguishes them from typical computer science programs.

Based on this,
proposed SE curricula combine material
from adjacent education areas
to provide SE practitioners with well-rounded knowledge.
In the seventies, Freeman \etal~\cite{10.5555/800253.807660}
presented the essential elements of SE education:
computer and management science topics,
communication skills, problem solving, 
design methodologies.
Later, a structure for graduate degree programs was introduced
by Freeman and Wasserman~\cite{10.5555/800099.803190},
balancing fundamental and specialized SE topics,
due to the difficulty of fitting all required
material in undergraduate programs.
Jensen \etal~\cite{10.1145/800130.804240}
combined elements from
computer science and engineering,
management science, communication and problem solving.
More recently,
Cowling~\cite{658297} presented a multi-dimensional model
for producing various curricula for computing and SE.

Researchers examine how university curricula
can better prepare software practitioners.
In the late nineties,
Lethbridge~\cite{lethbridge1998survey} surveyed software developers
to identify important SE and computer science topics
adapted from
the IEEE/ACM Software Engineering Body of Knowledge (SWEBOK)~\cite{SWEBOK14},
concluding that there is significant room for improvement
in what is taught to SE students.
The survey results influenced the design of the first SE program in Canada,
at the University of Ottawa.
Mathematics and engineering subjects were assessed as the least memorable,
as opposed to software topics which were found the most memorable.
Certain typically compulsory subjects including
differential equations and numerical methods
were found to be of low importance,
and were turned to optional in the particular program.
The author suggests that
more emphasis should be given to the components of
testing and quality assurance,
requirements analysis,
project management, and
configuration management---these are usually optional or receive less attention.
The study was repeated in 2000, again concluding that
traditional engineering courses
such as linear algebra, differential equations, and calculus
are overtaught, while more demanded SE courses
are inadequately covered~\cite{841783}.
Less weight should be placed on traditional courses,
and more on people skills, software processes,
human-computer interaction.

Studies recommend tighter collaboration between academia and industry
to better prepare trainees to meet industrial demands.
To train students and improve their ability of real-world problem solving,
in the mid 2000s,
Reichlmay~\cite{10.1145/1137842.1137848} encouraged closer
collaboration between academia and the SE professional industry.
Bollin \etal~\cite{6245008} proposed using a simulation program
to help participants familiarize with regular tasks
of professional software engineers.
A team-alternation methodology was suggested
by Walker and Slotterbeck~\cite{10.5555/775742.775765}
to assist undergraduates in dealing with complications occurring
when working for years on the same SE project.

Industrial needs for SE professionals
are continuously increasing.
In 2012,
Moreno \etal~\cite{moreno2012balancing}
examined the IEEE/ACM 2004 Software Engineering~\cite{10.1145/1062455.1062571}
and 2009 Graduate Software Engineering~\cite{4812713} curriculum recommendations
to evaluate whether undergraduate courses
effectively cover the market's growing needs.
They found gaps in the de facto courses,
particularly in the ones that tackle business-oriented topics
(\eg IT Business Consultancy)
as well as soft skills of software practitioners
(\eg negotiation skills).

This study aims to extend existing work on gaps
between industrial needs and software practitioners' education
using the Software Engineering Education Knowledge (SEEK)
of the SE2014~\cite{ABHO14}.
Building on top of already-identified shortcomings
of SE curriculum models,
this work investigates the industrial needs of developers,
assesses their satisfaction by the SEEK,
and suggests potential directions for improvement.
Future SEEK models and the SE education in general
could adapt their content
according to the findings of this work.
Refined curricula could provide future developers
with a stronger SE background
adapted to the industrial needs,
improving their professional careers and absorption.

Insights are provided on the satisfaction of developers' needs,
as expressed through Wikipedia articles referenced in Stack Overflow (SO) posts.
Wikipedia is frequently used for academic and professional activities,
with technology-related articles being among the most commonly visited~\cite{SLWZ17},
while SO is considered a popular programmer-oriented Q\&A forum.
According to the 2021 SO developer survey~\cite{stackoverflow},
70\% of respondents have at least a Bachelor's degree (or equivalent)
and 70\% are professional developers.
These are mainly occupied in
full-stack (50\%),
back-end (44\%),
front-end (27\%),
desktop (17\%),
mobile development (15\%),
and DevOps (11\%) positions.
The majority (63\%) have up to ten years of professional coding experience,
meaning they have never worked in a world without SO.
Moreover, 80\% of respondents visit SO weekly,
and 55\% of them visit daily.
As a result,
a considerable number of SO users are engaged with SE,
with the majority being new in the field.
Based on the insights derived from Wikipedia and SO,
the needs' satisfaction is assessed through the following research question.
\begin{description}
\item[RQ1] \emph{To what extent are developers' needs,
in terms of Wikipedia articles referenced in SO posts,
covered by the SEEK knowledge units?}
\end{description}
To examine
whether article popularity can indicate less satisfied needs,
the relationship
between SEEK coverage and Wikipedia article popularity is investigated
through the following research question,
in terms of citation frequency and votes in SO posts.
\begin{description}
\item[RQ2] \emph{How does the popularity of Wikipedia articles relate to their SEEK coverage?}
\end{description}
To facilitate future curriculum reports, and
limit the gap between industrial needs and education,
computing areas underrepresented in the SEEK are listed
through the research question:
\begin{description}
\item[RQ3] \emph{What areas of computing knowledge can be better covered by the SEEK knowledge units?}
\end{description}
Finally,
qualitative insights about programmers' behavior on SO are provided
through the research question:
\begin{description}
\item[RQ4] \emph{Why are Wikipedia articles covered by the SEEK knowledge units cited on SO?}
\end{description}

To validate the study findings,
a survey was conducted on SE practitioners.
Following published recommendations~\cite{IHG12},
the code\footnote{\url{https://doi.org/10.5281/zenodo.4251099}} and
data (analyzed Wikipedia articles and SO posts,
survey questionnaire and responses)\footnote{\url{https://doi.org/10.5281/zenodo.5055546}}
associated with this endeavor are openly available online,
and can be used for replication purposes,
or to perform further empirical studies.

\section{Related Work}
\label{sec:related}
The authors of this study consider as related work
literature reviews and surveys investigating the gap between
SE education and industrial needs, and
empirical studies evaluating the role of Wikipedia and SO
as learning resources for SE students.

Garousi \etal~\cite{8664487} conducted a systematic literature review
on the gap
between SE education and industrial needs.
Topics were classified
according to their importance and gap observed in the relevant literature.
Many topics were categorized both as
\emph{high-importance} and \emph{high-gap},
signifying potential deficiencies
in existing curricula.
Software requirements, quality, design,
testing, and project management
are highly important domains,
but insufficiently covered by most SE
undergraduate programs.
Despite the considerable demand for computer fundamentals,
little weight is placed by existing curricula,
while many programmers feel they have been educated more on
engineering foundations and mathematics
than on software topics.

The role of Wikipedia
as an opportunistic learning resource for prospective software engineers
is controversial.
While some consider the website useful for students
under the right circumstances~\cite{doi:10.1080/13614533.2012.740439},
others disapprove its use for learning new computing concepts,
arguing that novice learners should turn to online programmer communities
to satisfy their learning needs~\cite{10.1145/3328778.3366832}.
However, these communities often lead back to Wikipedia.
For instance,
Wikipedia references
in the August 2012 SO data dump~\cite{MSRChallenge2013}
account for about 5\% of total references~\cite{gomez2013study},
signifying the important contribution
of Wikipedia to online programmer communities.

\section{Methods}
\label{sec:methods}

\subsection{Collection of Wikipedia articles from SO posts}
\label{sec:collection}

Wikipedia articles referenced in SO posts (questions and answers)
were systematically collected
from the December 2019 SO data dump.\footnote{\url{https://archive.org/details/stackexchange2019-12-02}}
The dump contains 18\,597\,996 questions and 28\,248\,207 answers
posted on the website up to the aforementioned date,
along with metadata (\eg votes).
URLs from all Wikipedia language editions (not only the English subdomain)
were retrieved through automatic search.
Additionally, all different URL formats\footnote{\url{https://en.wikipedia.org/wiki/Help:URL}}
and non-clickable references (\eg in images or text quotes)
were taken into account.
Each post may also contain multiple references.
In total,
474\,798 references were retrieved, with 32\,332 of them being distinct.

Article titles were isolated from URLs, and
their format was canonicalized.
Format deficiencies concerned junk characters (\eg hash signs)
and missing punctuation (\eg closing brackets).
Multiple titles might redirect to the same Wikipedia article (\eg \emph{URL}, \emph{web address}).
These titles were resolved through the MediaWiki API\footnote{\url{https://www.mediawiki.org/wiki/API:Main_page}}
by mapping and renaming them to their redirection article.
To assess popularity,
titles were complemented with their appearance frequencies and cumulative post scores
(\ie total upvotes minus downvotes).

\subsection{RQ1,2: Mapping of top-referenced Wikipedia articles to the SEEK knowledge units}
\label{sec:association}
The most cited Wikipedia articles were manually assigned
to the SE2014 SEEK knowledge units,
which ``provide the foundation
for the design, implementation, and delivery of the educational units
that make up a SE curriculum''~\cite{ABHO14}.
Thus, SEEK was preferred over a more general computing education guide.
In addition,
the 2020 ACM/IEEE Computing Curricula report \cite{CIPW21},
which provides undergraduate curriculum guidelines for the computing education,
recognizes the SE2014 as the most suitable and up-to-date guide
for the SE discipline.

Five hundred Wikipedia articles,
accounting for the top 1\% of total references,
were manually assigned to the SEEK's 37 knowledge units.
This percentage was selected
by balancing the need for an adequate article coverage
with the available time margin and human resources.
In the end,
assigned units were aggregated to their general knowledge areas
to compute further statistics.

To ensure consistency of this manual process,
recommended guidelines suggested in the work
of Brereton \etal~\cite{BKBT07} were followed.
Two data extraction approaches were combined:
the use of two reviewers
performing individually the data extraction process
and discussing their disagreements; and
the use of a data extractor and a data checker.

Following these,
the first two authors of this paper
individually assessed (in a spreadsheet)
the extent of coverage of each Wikipedia article
and assigned it to a knowledge unit (if any coverage was noted).
The coverage range included the following options:
\emph{fully covered};
\emph{partially covered};
\emph{off topic}---the article's topic is not related to any SE field;
\emph{general knowledge};
\emph{extremely specialized}; and
\emph{not covered}---the topic should be covered but is not
(and is also not off topic, general knowledge, or extremely specialized).
To amplify human judgment,
after observing patterns between articles and knowledge units,
a set of assertions
was defined by all authors at the beginning of the procedure,
reinforced during the mapping process---these
are the following.
\begin{itemize}
\item Article topics included
in a SEEK knowledge unit's description are fully mapped to it.
\item Articles including general information
irrelevant to any SEEK unit are \emph{off topic}.
\item Articles on web development and web design
are \emph{not covered} by the SEEK.
\item Articles about encoding are \emph{extremely specialized},
because they are too specific with regard to the SEEK units.
\item Articles analyzing fundamental software design patterns are mapped
to \emph{Design Strategies} (DES.str),
while articles on more advanced pattern implementations
are mapped to \emph{Detailed Design} (DES.dd).
Articles listing fundamental software design principles
are mapped to \emph{Design Concepts} (DES.con).
\item Articles delving into database design are mapped
to \emph{Detailed Design} (DES.dd),
while articles about introductory database design concepts
are mapped to \emph{Computer Science Foundations} (CMP.cf).
\item Articles on established algorithms
are fully mapped to \emph{Computer Science Foundations} (CMP.cf),
while those referring to more specialized algorithms
are partially mapped to the same unit.
\end{itemize}

An example of a not covered article is
\emph{ISO-8601},\footnote{\url{https://en.wikipedia.org/wiki/ISO_8601}}
involving a widely used international standard 
for date and time representation 
that reduces program faults and complexity.
Although the topic could be encountered 
by a software engineer, 
no SEEK unit appears to cover it.
Other not covered article examples are
\emph{Cross-origin resource sharing},
\emph{JSONP},
\emph{Ajax (programming)},
\emph{Post/Redirect/Get},
\emph{Comet (programming)}.

A partially covered article is
\emph{Android (Operating System)}\footnote{\url{https://en.wikipedia.org/wiki/Android_(operating_system)}},
involving Android’s 
history, features, hardware, development, security, licensing, and legal issues. 
Although the topic could be covered by
\emph{Operating System Basics} (CMP.cf.9)
of \emph{Computer Science Foundations} (CMP.cf), 
it only provides an introduction without studying the topic adequately,
thus it was partially mapped to the unit. 
Other partially covered articles are 
\emph{UTF-8},
\emph{Trie},
\emph{Endianness},
\emph{Levenshtein distance},
\emph{JSON}.

Upon completing the assessment,
the two authors discussed and partially resolved their disagreements,
documenting their opinion.
Through this process,
5\% of assignments remained conflicting.
These were resolved by the paper's last author,
again by selecting from all options,
drawing on the first two authors' assertions and documented comments
as well as his three-decades' experience in SE.
In the end,
25\% of pending disagreements were resolved in favor of the first author,
and 38\% in favor of the second.
The overall agreement rate between the last and the first two authors was 63\%.

\subsection{RQ3: Classification of partially and not covered Wikipedia articles}
\label{sec:classification}
Wikipedia articles characterized as partially or not covered
through the process described in Section~\ref{sec:association}
were manually classified using the
2012 ACM Computing Classification System (CCS)~\cite{Rou12}.
The first two authors of this paper together mapped each article by hand
to a knowledge area and subarea of the first and second level categories
of the system, correspondingly,
taking into account the lower level contents of the areas.
Again, any reservations that occurred were resolved by the paper's last author.

\subsection{RQ4: Manual coding of SO posts with references to covered Wikipedia articles}
\label{sec:coding}
To examine why articles covered by the SEEK knowledge units are cited
on SO by developers,
the first two authors of this paper applied manual coding~\cite{CS90}
to a sample of posts referencing covered Wikipedia articles.
The sample size of a population of 198\,365 posts was calculated at around 384
using Cochran's~\cite{Coc77} sample size and correction formula
depicted in Eq.~\ref{eqn:cochrans_equation},
which provides a representative sample for proportions of large populations.
\begin{equation}
\label{eqn:cochrans_equation}
n_0 = \frac{z^2pq}{e^2},
\end{equation}
where $n_0$ is the required sample size, 
$e$ the desired margin of error (5\%), 
$p$ the proportion of population (0.5---maximum variability),
$q = 1 - p$,
and $z$ is found in the $Z$ table.

Posts were split in two, 
and each author individually applied codes to a half (in a shared online spreadsheet). 
The authors read each randomly selected post (question/answer) 
and produced codes related to the post's content. 
Posts with similar content and area of question were assigned the same code. 
At least one and up to five codes were applied to each post. 
Next, the authors discussed and grouped together 
conceptually-related codes 
by generalizing or specializing them. 
In the end,
thirteen aggregated codes (listed in Section~\ref{sec:rq4}) occurred.

As a result,
a post might eventually be associated with more than one general code.
For polytomous or continuous variables,
Israel~\cite{Isr92} in his study on determining sample size
suggests using the formula for sample size for the mean,
which is similar to Cochran's formula for the proportion
(Eq.~\ref{eqn:cochrans_equation}),
except for the measure of variability;
the formula for the mean employs $\sigma^2$,
which is the variance of an attribute in the population,
instead of $(p \times q)$.
However,
using the level of maximum variability ($p=0.5$)
in the calculation of the sample size for the proportion
generally produces a more conservative sample size (\ie a larger one)
than the formula for the mean.
In this study the level of maximum variability was employed,
thus the sample size is considered more conservative
than the one that would be calculated by the sample size for the mean.

\subsection{Validation Survey}
\label{sec:survey}
A survey was conducted on SE practitioners to validate the study findings.
Following Kitchenham and Pfleeger's survey guidelines~\cite{10.1145/566493.566495},
the authors adopted a descriptive survey design
by performing a cross-sectional, case control study
(\ie participants were surveyed about their past experiences
at a particular fixed point in time),
which is typical of surveys in SE.
The survey questionnaire consisted of multiple-choice, open-ended,
and Likert-scale questions~\cite{10.1145/511152.511155}.
Documentation was included about the survey objective and study context,
an optional question for comments on the questionnaire,
and an option for participants to complete their e-mail address
and receive a report with the survey results.
The complete questionnaire and responses are included in the dataset
(Section~\ref{sec:intro}).

To evaluate the questionnaire, 
a pilot survey was conducted~\cite{10.1145/638574.638580}
on the members of the authors' laboratory.
The questionnaire was improved based on four members' feedback,
and the revised version was e-mailed to 23 SE practitioners
from eight organizations:
Deloitte Greece;
Deloitte Poland;
Deloitte Spain;
SICOA Consulting;
Quintessential;
GRNET;
National Bank of Greece;
Workable.
The final survey ran from June 23rd to 29th, 2021, and
seven responses were received (30\% response rate).
Four respondents work in software development positions,
two are business analysts,
and one is a software consultant.
Five have graduated 
from a higher education institution
with a SE-related major,
one is currently a university student majoring in telecommunications, and 
one holds a high school diploma.
All respondents have been practicing SE as professionals.

\section{Results}
\label{sec:results}

\subsection{RQ1: To what extent are developers' needs, in terms of Wikipedia articles referenced in SO posts, covered by the SEEK knowledge units?}
\label{sec:rq1}
The 500 manually analyzed Wikipedia articles (Section~\ref{sec:association})
resulted in
306 (61.2\%) fully and 88 (17.6\%) partially covered,
29 (5.8\%) not covered,
13 (2.6\%) off-topic,
17 (3.4\%) general-knowledge, and
47 (9.4\%) extremely specialized cases.
Ultimately, 79\% of articles discussed on SO are covered
by the SEEK.

The coverage of programmers' needs was assessed
based on the fully and partially covered Wikipedia articles.
The knowledge unit of \emph{Computer Science Foundations} (CMP.cf), which
focuses on the provision of basic SE knowledge,
was mapped to the majority (34.6\%) of articles.
Succeeding units are \emph{Detailed Design} (DES.dd---8.2\%),
\emph{Mathematical Foundations} (FND.mf---7.2\%),
\emph{Construction Technologies} (CMP.ct---5.8\%),
\emph{Architectural Design} (DES.ar---3.2\%),
\emph{Security Fundamentals} (SEC.sfd---3\%),
\emph{Design Strategies} (DES.str---2.6\%),
\emph{Construction Tools} (CMP.tl---2.2\%),
\emph{Design Concepts} (DES.con---2.2\%),
\emph{Problem Analysis and Reporting} (VAV.par---1.4\%),
\emph{Computer and Network Security} (SEC.net---1.4\%),
\emph{Developing Secure Software} (SEC.dev---1.4\%),
\emph{Engineering Foundations for Software} (FND.ef---1.2\%),
\emph{Human-Computer Interaction Design} (DES.hci---1.2\%),
\emph{Software Configuration Management} (PRO.cm---1\%),
\emph{Types of Models} (MAA.tm---0.6\%),
\emph{Testing} (VAV.tst---0.6\%),
\emph{Evolution Process and Activities} (PRO.evo---0.6\%),
\emph{Requirements Fundamentals} (REQ.rfd---0.2\%),
\emph{Design Evaluation} (DES.ev---0.2\%).

The remaining units were not mapped to any article.
These are
\emph{Engineering Economics for Software (FND.ec)},
\emph{Group Dynamics and Psychology} (PRF.psy),
\emph{Communication Skills} (PRF.com),
\emph{Professionalism} (PRF.pr),
\emph{Modeling Foundations} (MAA.md),
\emph{Analysis Fundamentals} (MAA.af),
\emph{Eliciting Requirements} (REQ.er),
\emph{Requirements Specification and Documentation} (REQ.rsd),
\emph{Requirements Validation} (REQ.rv),
\emph{Verification and Validation Terminology and Foundations} (VAV.fnd),
\emph{Reviews and Static Analysis} (VAV.rev),
\emph{Process Concepts} (PRO.con),
\emph{Process Implementation} (PRO.imp),
\emph{Project Planning and Tracking} (PRO.pp),
\emph{Software Quality Concepts Culture} (QUA.cc),
\emph{Process Assurance} (QUA.pca), and
\emph{Product Assurance} (QUA.pda).

\subsection{RQ2: How does the popularity of Wikipedia articles relate to their SEEK coverage?}
\label{sec:rq2}
The relation between article popularity and coverage
was examined by rating Wikipedia articles
according to their frequency in SO posts
and the associated posts' cumulative score (Section~\ref{sec:collection}).

\begin{figure}[!t]
\centering
\includegraphics[width=0.48\textwidth, keepaspectratio]{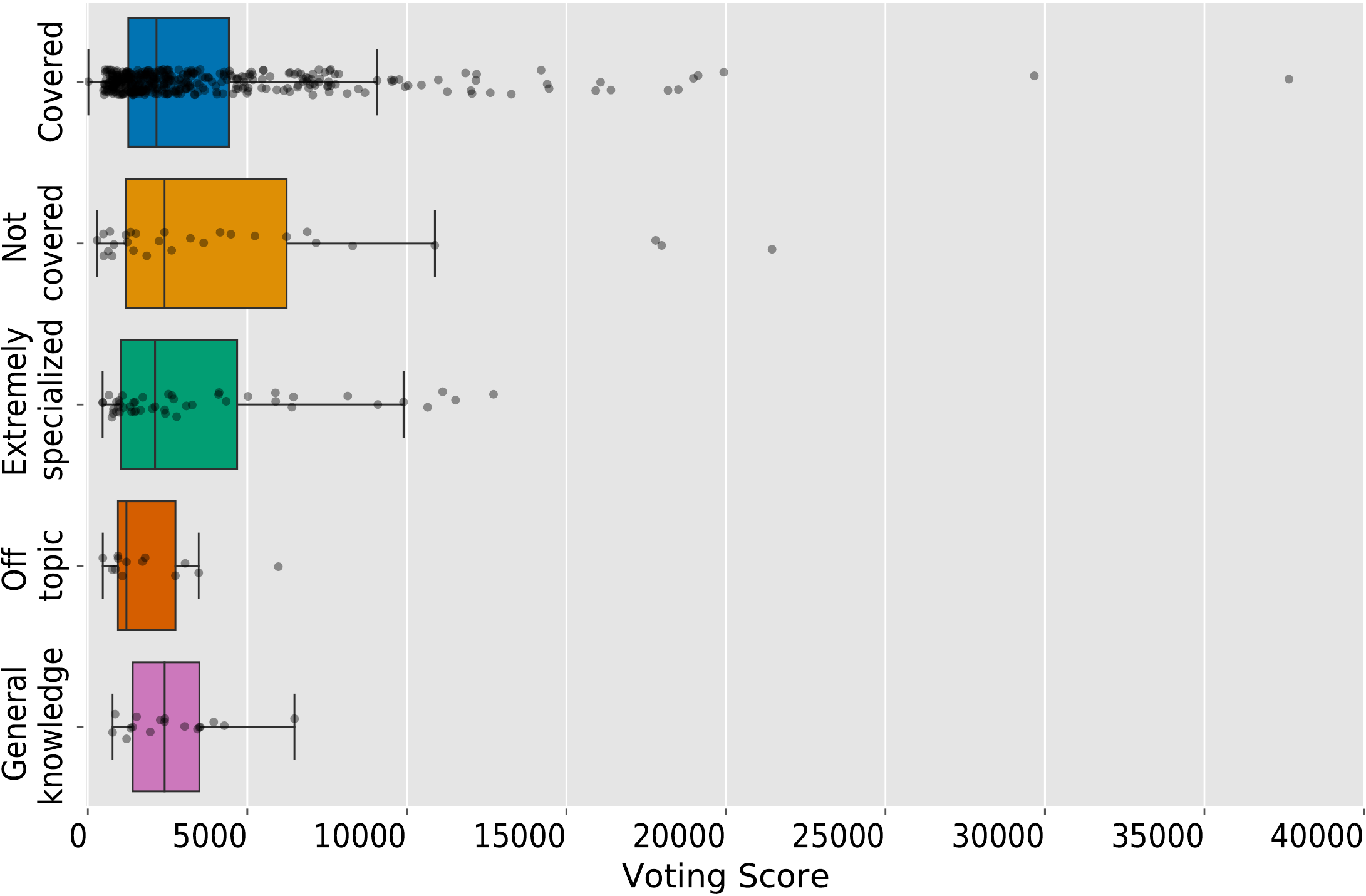}
\caption{
Article score per category of coverage.
Covered, not covered, and extremely specialized articles have similar score distributions,
while off-topic and general-knowledge articles are concentrated in the lower score areas.
}
\label{fig:score}
\end{figure}

Figure~\ref{fig:score} displays
for each coverage category
the cumulative post voting scores of the corresponding articles.
The majority of covered articles are characterized by low-to-moderate popularity,
but there is also a noteworthy number of articles that are extremely popular.
However,
the scores of not covered articles seem to vary widely,
ranging from low to high popularity,
with only few very popular cases.
Extremely specialized articles depict little differences,
with a slightly reduced variance,
while off-topic and general-knowledge articles mainly display low-to-moderate popularity.

\begin{figure}[!t]
\centering
\includegraphics[width=0.48\textwidth, keepaspectratio]{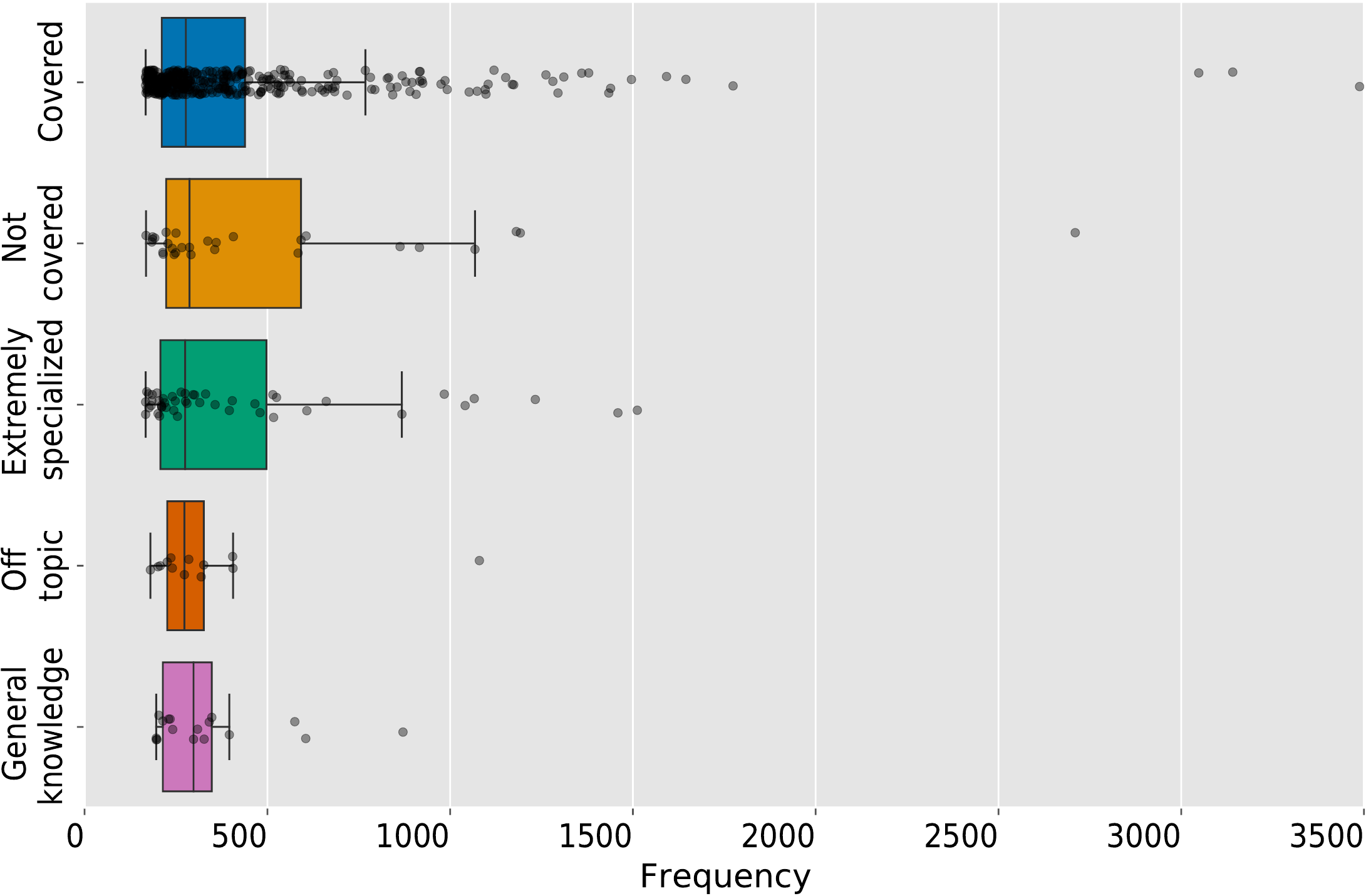}
\caption{
Article frequency per category of coverage.
Off-topic and general-knowledge articles have reduced frequency variance,
while the remaining categories' distribution resembles that of Fig.~\ref{fig:score}.
}
\label{fig:frequency}
\end{figure}

Equivalently,
Fig.~\ref{fig:frequency} demonstrates article frequency per category of coverage.
Minor differences to Fig.~\ref{fig:score} are observed
in the distribution of the articles for each category,
with general-knowledge articles showcasing noticeably reduced variance.
Survey respondents assessed the 20 most cited
not covered and partially covered articles as of low importance,
aligning with the above observations.

\subsection{RQ3: What areas of computing knowledge can be better covered by the SEEK knowledge units?}
\label{sec:rq3}
Through the process described in Section~\ref{sec:classification},
Wikipedia articles characterized as partially or not covered 
were assigned to the 2012 ACM CCS~\cite{Rou12}
first and second level knowledge areas.
Table~\ref{tab:notcovered} presents the frequency
of areas and subareas of not covered articles.
The majority (55\%) fall within the area
of \emph{Information systems},
particularly in the subarea of \emph{World Wide Web} (45\%),
followed by \emph{Computing methodologies} (20\%)
and \emph{Software and its engineering} (10\%).

Table~\ref{tab:partiallycovered} introduces the knowledge areas
of partially covered articles.
The majority (49\%) correspond to the area of
\emph{Software and its engineering},
specifically to the subarea of \emph{Software notations and tools} (43\%).
Subsequent areas concern \emph{Information systems} (28\%) and
\emph{Theory of computation} (14\%).

To further investigate what areas can be better covered by the SEEK,
survey participants assessed the SEEK knowledge units'
importance and recommended lecture hours.
\emph{Computer Science Foundations} (CMP.cf) and 
\emph{Testing} (VAV.tst) 
were rated as the most important SEEK units,
while less important units are
\emph{Mathematical Foundations} (FND.mf),
\emph{Professionalism} (PRF.pr),
\emph{Modeling Foundations} (MAA.md),
\emph{Software Configuration Management} (PRO.cm),
\emph{Evolution Process and Activities} (PRO.evo).
The findings indicate that
more lecture hours should be devoted
to \emph{Security Fundamentals} (SEC.sfd),
\emph{Computer and Network Security} (SEC.net), and
\emph{Developing Secure Software} (SEC.dev)---these all
belong to the \emph{Security} (SEC) knowledge area.
In addition,
more hours should be allocated to
\emph{Architectural Design} (DES.ar),
\emph{Design Concepts} (DES.con), and
\emph{Human-Computer Interaction Design} (DES.hci).
The SEEK could be enriched with
problem solving approaches in less technical situations
(\eg efficiently handling setbacks before an app launch),
guidelines for high-quality code writing,
software life cycle models such as Agile 
and its methods (\eg Scrum),
as well as
individual and professional growth topics (\eg interview simulations).

Survey participants also assessed a set of topics
that were found to be partially or not covered by the SEEK.
From these,
they consider algorithms as the most important topic,
followed by the World Wide Web, 
specialized software tools and systems (\eg database management systems), and
computer graphics.
Respondents suggest that
SE curricula should be updated regularly,
every one to four years.

\begin{table}[!t]
\renewcommand{\arraystretch}{1.2}
\caption{2012 ACM CCS~\cite{Rou12} Areas of Not Covered Articles}
\label{tab:notcovered}
\centering
\resizebox{0.48\textwidth}{!}{
\begin{tabular}{l l r r}
\hline
Area				& Subarea						& Articles	& (\%)\\
\hline
Information systems		& World Wide Web                            		& 13		& 45\\
				& Information retrieval					& 2		& 7\\
				& Data management systems				& 1		& 3\\
Computing methodologies		& Computer graphics                         		& 4		& 14\\
				& \thead{Parallel computing \\ methodologies}		& \ar{1}	& \ar{3}\\
				& \thead{Symbolic and algebraic \\ algorithms}       	& \ar{1}	& \ar{3}\\
Software and its engineering	& Software notations and tools              		& 2		& 7\\
				& \thead{Software creation and \\ management}        	& \ar{1}	& \ar{3}\\
\al{Applied computing}		& \thead{Document management and \\ text processing} 	& \ar{1}	& \ar{3}\\
\al{General and reference}	& \thead{Cross-computing tools and \\ techniques}	& \ar{1}	& \ar{3}\\
Mathematics of computing	& Mathematical analysis                     		& 1		& 3\\
Networks			& Network protocols                         		& 1		& 3\\
\hline
\end{tabular}}
\end{table}

\begin{table}[!t]
\renewcommand{\arraystretch}{1.2}
\caption{2012 ACM CCS~\cite{Rou12} Areas of Partially Covered Articles}
\label{tab:partiallycovered}
\centering
\resizebox{0.48\textwidth}{!}{
\begin{tabular}{l l r r}
\hline
Area				& Subarea						& Articles	& (\%)\\
\hline
Software and its engineering	& Software notations and tools            & 38		& 43\\
				& \thead{Software organization and \\ properties}	& \ar{4}	& \ar{5}\\
				& \thead{Software creation and \\ management}		& \ar{1}	& \ar{1}\\
Information systems			& Data management systems              & 12		& 14\\
				& World Wide Web						& 10		& 11\\
				& Information retrieval					& 3		& 3\\
\al{Theory of computation}		& \thead{Design and analysis of \\ algorithms}     & \ar{11}		& \ar{13}\\
				& Semantics and reasoning					& 1		& 1\\
Mathematics of computing		& Discrete mathematics                      & 1		& 1\\
				& Mathematical analysis					& 1		& 1\\
				& Computer graphics						& 1		& 1\\
Computing methodologies		& Artificial intelligence                        & 1		& 1\\
				& \thead{Concurrent computer \\ methodologies}		& \ar{1}	& \ar{1}\\
Security and privacy			& Cryptography                     		& 2		& 2\\
General and reference		& Document types                         	& 1		& 1\\
\hline
\end{tabular}}
\end{table}

\subsection{RQ4: Why are Wikipedia articles covered by the SEEK knowledge units cited on SO?}
\label{sec:rq4}
Through the process described in Section~\ref{sec:coding}
a variety of reasons emerged
justifying why articles covered by the SEEK knowledge units
are cited in SO questions and answers.
The aggregated codes assigned to the examined posts were
\emph{programming language-specific explanation/advice} (24\%), 
\emph{advice for system/software design} (21\%),
\emph{code review/debugging} (20\%),
\emph{computer science/programming concept explanation} (11\%),
\emph{algorithm recommendation/optimization} (9\%),
\emph{tool recommendation} (4\%),
\emph{general discussion} (3\%),
\emph{exercise help/provide solution} (2\%),
\emph{data structure recommendation} (2\%),
\emph{testing techniques recommendation} (2\%),
\emph{mathematical concept explanation} (1\%),
\emph{learning material recommendation} (1\%),
\emph{detailed explanation of posted code} (1\%). 
One post had been deleted from SO,
and was assigned the code \emph{N/A}.

Survey participants mainly agree with the above reasons.
Furthermore,
they use Wikipedia articles to validate their arguments 
or understand theoretical concepts,
while the majority stress that SO can also be used
for purposes unrelated to technical issues.

\section{Discussion and Implications}
\label{sec:discussion}

The majority of top-referenced Wikipedia articles
concerning SE topics are covered by the SEEK.
A striking example is the number of articles mapped
to \emph{Computer Science Foundations} (CMP.cf).
A reason for the unit's dominance is
the wide range of knowledge it embodies
and the considerably more lecture hours it requires.
The SEEK emphasizes the importance of
the foundational computer science concepts, 
serving as a springboard to more advanced concepts, and
this is also validated by survey respondents.
The associated need described in Section~\ref{sec:related} seems to be met.

Many articles regarding mathematical topics were observed
and mapped to \emph{Mathematical Foundations} (FND.mf).
This contradicts arguments that
SE courses should focus less on mathematical concepts.
For example, Lethbridge
concludes that mathematical topics 
including calculus, differential equations, and linear algebra 
have a negative educational gap
(\ie the knowledge learned is greater than their importance),
hence less emphasis could be placed on them~\cite{841783}.
However,
the manual analysis of posts referencing covered Wikipedia articles
(Section~\ref{sec:rq4}) revealed that
few practitioners seek explanations about mathematical concepts,
suggesting a lower demand for them. 
Explanations involve geometry and combinatorics problems
applied to specialized programming issues.
Mathematical foundations in SE curricula
may be sufficient for an average developer,
thus it might be worth reducing weight on more advanced concepts
to allow space for less covered but highly demanded topics.

\begin{implication}
The SEEK seems to effectively provide developers
with the mathematical foundations they need.
Further analysis of SE curricula could indicate
whether reducing some weight on more advanced mathematical concepts,
releasing space for less covered but highly demanded topics,
would have a positive impact on developers.
\end{implication}

Survey respondents perceived as extremely important
the area of software design.
Although various articles were mapped to
units of the knowledge area \emph{Software Design} (DES),
survey participants recommended allocating more hours to
\emph{Architectural Design} (DES.ar),
\emph{Design Concepts} (DES.con), and
\emph{Human-Computer Interaction Design} (DES.hci).
This is reinforced by the fact that
one of the most common reasons
for citing covered Wikipedia articles in SO posts concerns
developers seeking advice for software design and development,
such as best implementation practices.

\begin{implication}
Further investigation could evince the impact
of extending the content of the SEEK on practical and specialized approaches
related to software design concepts,
architectural design, and human-computer interaction design.
\end{implication}

Two other equally important areas for practitioners,
according to survey participants,
are
software testing and security.
They recommended extending the lecture hours
of all units included in the area of \emph{Security} (SEC).
Although few Wikipedia articles were related to \emph{Testing} (VAV.tst),
possibly because such topics may not be
as popular in online programmer communities,
still,
the incorporation of more testing material in the SEEK
could equip students with useful skills 
for their professional careers.

\begin{implication}
Further analysis could reflect the value
of including more material and allocating more lecture hours
to the areas of software testing and security in the SEEK.
\end{implication}

Articles only partially covered by the SEEK
are associated with
specialized software tools and notations, database management systems,
and algorithms.
Focusing on SE education,
the SEEK manages to cover the most relevant articles.
However, more specialized programming tools and notations
discussed on SO
are an exception to the rule,
either because they are too specialized 
to be fully covered by SE education,
or they are no longer widely used,
hence are nowadays less covered by curriculum models.
Additional analysis could indicate whether
these gaps should be covered by the SEEK or industrial training.

Additionally,
articles related to World Wide Web principles and tools,
information retrieval, and computer graphics
are less likely to be covered by the SEEK units---the SEEK seems
to cover these areas rather superficially.
Particularly,
the area of computer graphics may be only partially covered
because it can be considered less related to SE.
Nevertheless, the remaining areas,
despite being widely discussed by the community,
are barely mentioned in introductory units,
revealing potential unsatisfied needs of programmers.

\begin{implication}
The SEEK could emphasize more information systems,
particularly,
information retrieval methods, 
World Wide Web principles and tools, and computer graphics.
\end{implication}

The majority of knowledge units
not associated with any Wikipedia reference on SO
are related to SE concepts with
a strong presence of the human factor.
These fall within the knowledge areas of
\emph{Professional Practice} (PRF),
\emph{Requirements Analysis and Specification} (REQ),
\emph{Software Process} (PRO), and
\emph{Software Quality} (QUA),
aligning with the deficiencies outlined in Section~\ref{sec:related}.

In addition,
a weak relation was found between article popularity and coverage,
regardless of the metric (frequency, voting score) used (Section~\ref{sec:rq3}).
The distribution of both covered and not covered articles is almost identical.
The notably reduced number
of not covered observations 
may partially explain the lack
of extremely popular not covered articles,
compared to the covered ones.
Although most articles are covered by the SEEK,
the ratio between covered and not covered articles remains unchanged,
regardless of their popularity.
However,
the popularity of off-topic and general-knowledge articles is remarkably lower,
especially in terms of score.
Perhaps articles that are either unrelated to computer science, or
are considered conventional SE concepts,
receive less attention
and are rarely mentioned in online programmer communities.
However, survey respondents 
contradict that SO is solely used for technical issues.

Various reasons for citing covered Wikipedia articles on SO
were identified during the manual coding process 
described in Section \ref{sec:coding}.
Programmers often demand recommendations or explanations regarding
SE components such as
algorithms and data structures,
programming languages,
frameworks,
testing techniques, and
computer science concepts; this highlights
the importance for curriculum models to provide deep understanding
of computer science and SE fundamentals,
aligning with the survey findings.
Although programmers typically
introduce code snippets to stress their point and
provide descriptive examples,
there are also many instances where they solely request
code review and debugging assistance,
as they struggle to detect
errors and flaws in their code.
Furthermore,
developers look for learning material recommendations and
tool suggestions
to broaden their horizons and
expand their professional skills.
This demonstrates their willingness to
enhance their abilities in the SE field,
embracing the idea of autodidacticism.
They frequently need assistance in exercises and problems
related to their studies or work,
revealing a difficulty in applying
theoretical knowledge to practice.
Finally,
developers do not always seek solutions to issues,
but aim to initiate general discussions.

\section{Threats to Validity}
\label{sec:threats}
The study's external validity in terms of generalizability suffers
by studying the SE2014 SEEK model
rather than a real-world curriculum for SE undergraduate studies.
However, the SEEK has been thoroughly composed
based on input from the SE community
through a survey, several workshops and informal discussion sessions,
and consists of the essential material
for the development of any curriculum in the field~\cite{ABHO14}.
In addition,
results mostly target SE education,
but could be generalized to other computing fields
through study replication using the corresponding ACM curriculum
reports.\footnote{\url{https://www.acm.org/education/curricula-recommendations}}

Threats to the study's internal validity stem from the steps
during which manual processes involving subjective judgment were followed:
the categorization of Wikipedia articles based on the SEEK knowledge units,
the classification of partially and not covered articles using the 2012 ACM CCS,
and the manual coding of SO posts.
The risk stemming from the latter process is related
to the loss of accuracy of the original post
due to an increased level of categorization;
this threat was reduced
by assigning multiple codes to each post.
The same process is affected by the restricted number of analyzed posts
and the selected sampling technique,
which may not be the most appropriate one for the SO population.
(For example,
SO users must earn reputation points to perform certain acts,
while users may exist that
interact with large numbers of posts
to gain reputation for purposes unrelated to SO.)
This limitation could be improved
by increasing the sample size in replication studies
or by employing different sampling techniques
that account for the complexity of the SO population.
The trustworthiness of the manual processes was enhanced
through the use of multiple raters,
and by grounding them on established research methods.
However,
validity risks derived from manual processes
requiring human judgment cannot be completely eliminated~\cite{PVK15}.

Threats to the internal validity also result
from the evaluation of programmers' needs using the SEEK, and
the survey process.
For the assessment of developers' needs using the SEEK (Section~\ref{sec:rq1}),
all knowledge units were considered equally likely
to appear on SO through Wikipedia article references.
However,
some units associated with fewer references could be less challenging
for developers, resulting in fewer posts,
thus one should not overinterpret under- or unrepresented SEEK units on SO.
Threats stemming from the survey process include
the small sample size which prevents generalizability of results,
the social desirability bias~\cite{furnham1986response},
which may have affected respondents' honesty,
and the question-order effect~\cite{sigelaman1981question},
which may have directed participants' answers.
To reduce the desirability bias,
participants were informed about the anonymity of their responses.
Although questions could have been shuffled to avoid the order effect,
they were sorted in a convenient manner
to assist respondents' comprehension.

\section{Conclusion}
\label{sec:conclusion}
This study aims to extend existing work on the gap between
software practitioners' education and industrial needs
by analyzing Wikipedia articles referenced in SO posts.
The analysis indicates that
common reasons developers reference Wikipedia articles on SO
include advice and best practices regarding software design,
explanations on computer science concepts, algorithms and data structures
as well as code review assistance.
Foundational computer science, mathematical, and design concepts preponderate
in covered articles,
while a noteworthy number of SEEK units related to the human factor found no match.
It could be the case that human-related topics
are not generally encountered on SO,
or the associated SEEK units
(\ie \emph{Group Dynamics and Psychology}---PRF.psy,
\emph{Communication Skills}---PRF.com,
\emph{Professionalism}---PRF.pr) 
are not discussed by SO users.
In addition, deficiencies were observed
in the areas of World Wide Web, computer graphics, and
specialized software tools.
Weak relation between article popularity (in terms of frequency and post score)
and coverage was observed.
A survey conducted on software practitioners
enhanced and reinforced the analysis findings.
It might be worth for future SEEK models
and the computing education in general
to dive deeper in the areas of
information systems, software testing, design, security,
and invest more on soft skills,
such as teamwork, collaboration, and project management.

\section*{Acknowledgment}
This work has received funding from the European Union's
Horizon 2020 research and innovation programme
under grant agreement No. 825328 (FASTEN project).

\ifCLASSOPTIONcaptionsoff
  \newpage
\fi

\bibliographystyle{IEEEtran}
\bibliography{LPKS21}

\end{document}